\documentclass[11pt]{article}

\usepackage{amssymb,amsmath,amsthm}

\theoremstyle{plain}
\newtheorem{prop}{Proposition}[section]
\newtheorem{thm}[prop]{Theorem}
\newtheorem{cor}[prop]{Corollary}
\newtheorem{lem}[prop]{Lemma}

\theoremstyle{remark}
\newtheorem{rem}{Remark}

\renewcommand{\theequation}{\thesection.\arabic{equation}}

\newcommand{\sst}{\scriptstyle}

\newcommand{\ds}{\displaystyle}
\newcommand{\vp}{\varepsilon}
\newcommand{\wt}{\widetilde}

\begin{document}

\title{Generalization of Grover's Algorithm to Multiobject\\
 Search in Quantum 
Computing, Part I: \\ Continuous Time and Discrete Time}

\author{Goong Chen$^{1,2}$ \and Stephen A.~Fulling$^{1}$ \and
 Jeesen Chen$^{3}$}

\date{}
\maketitle

\abstract{L. K.~Grover's search algorithm in quantum computing gives an
optimal,  quadratic speedup in the search for a single object in a large
unsorted  database. In this paper, we
generalize  Grover's algorithm in a Hilbert-space 
 framework for both continuous and discrete time cases that isolates its
geometrical essence to the case where  more than one
object satisfies the search criterion.}

 \vskip1truein

\vfil

\begin{itemize}
\item[1.] Department of Mathematics, Texas A\&M University, College Station, TX 
\ 77843-3368. E-mail:~gchen@math.tamu.edu and fulling@math.tamu.edu.
\item[2.] Supported in part by Texas A\&M University Interdisciplinary Research 
Grant IRI 99-22.
\item[3.] Department of Mathematical Sciences, University of Cincinnati,
Cincinnati, OH 45221-0025. E-mail:~chenjn@email.uc.edu.
\end{itemize}

\eject

\baselineskip = 18pt

\section{Introduction}\label{grov:sec1}

\indent

A quantum computer (QC) is envisaged as a collection of 2-state ``quantum 
bits''\negthinspace,
  or {\em qubits\/} (e.g., spin 1/2 particles). 
 Quantum computation does 
calculations on data densely coded in the 
 entangled states that are the hallmark of 
quantum mechanics, potentially yielding unprecedented parallelism in 
computation, as P.~Shor's work on factorization 
[\ref{shor1},~\ref{shor2}] proved in 1994. 
Two years later, L.~K.~Grover [\ref{grover1}]  showed that for an 
unsorted database with $N$ items in storage,
  it takes an average number of 
$\mathcal{O}(\sqrt N)$ searches to locate a single desired object 
by his quantum search algorithm. 
 If $N$ is a very large number,
 this is a significant quadratic 
speedup over the exhaustive search algorithm in a classical 
computer, which requires an average number of $\frac{N+1}2$ 
searches (see Remark~A.1 in Appendix). Even though Grover's algorithm is not
exponentially fast (as Shor's is), it has been argued that
 the wide range of its applicability compensates for this 
[\ref{BHT}]. 
Furthermore, the quantum speedup of the search algorithm
is {\em indisputable},
whereas for factoring the nonexistence of competitively
fast classical algorithms has not yet been proved
[\ref{Bih},~\ref{Bir}].

Grover's original papers [\ref{grover1}, \ref{grover2}] deal with 
 search for a single object. 
 In practical applications, typically more than one item  will 
satisfy the criterion used for searching.
In  the simplest generalization of Grover's algorithm,
 the number of ``good'' items is known in advance 
 (and greater than~$1$).
 Here we expound this generalization, 
along the lines of a treatment of the single-object 
 case by Farhi and 
Gutmann [\ref{FG}] that makes the 
 Hilbert-space geometry of the situation very clear.

 The success of Grover's algorithm and its multiobject generalization 
 is attributable to two main sources:
\begin{itemize}
\item[(i)] the notion of amplitude amplication; and
\item[(ii)] the dramatic reduction to invariant subspaces of low dimension for 
the unitary operators involved.
\end{itemize}
 Indeed, the second of these can be said to be responsible for the 
first:
 A proper geometrical formulation of the process shows that all the 
``action'' takes place within a {\em two-dimensional, real\/} 
subspace of the Hilbert space of quantum states.
 Since the state vectors are normalized, the state is confined to a 
one-dimensional unit circle and (if moved at all) initially has 
nowhere to go except toward the place where the amplitude for the 
sought-for state is maximized.
 This accounts for the robustness of Grover's algorithm --- that 
is,
 the fact that Grover's original choice of initial state and of the
 Walsh--Hadamard transformation can be replaced by (almost) any
 initial state and (almost) any unitary transformation
 [\ref{grover3}, \ref{jozsa},~\ref{BHT}].

The notion of amplitude amplification 
  was  emphasized  in the original works [\ref{grover1}, 
\ref{grover2}, \ref{grover3}] of Grover himself
 and in those of 
 Boyer, Brassard, H\o{}yer and Tapp [\ref{BBHT}] and 
 Brassard, H\o{}yer and Tapp~[\ref{BHT}].
(See also [\ref{Bih},~\ref{Bir}].)
 Dimensional reduction is prominent 
 in the papers by Farhi and Gutmann [\ref{FG}]  and Jozsa~[\ref{jozsa}]. 
We applied dimensional reduction to multiobject search
independently of references [\ref{BBHT}] and [\ref{BHT}] and later
learned that the same conclusions about multiobject search 
 (and more), in the {\em discrete time case}, had been 
obtained there in a different framework.
 
The rest of the paper is divided into three parts. In \S 2, we present the
continuous-time version of the multiobject search algorithm, and in \S 3
 the discrete-time version. In the Appendix, the computational complexity of
the classical random multiobject search algorithm is analyzed and some
relevant points on the literature are also made.

\section{Continuous Time Quantum Computing Algorithm for Multiobject 
Search}\label{grov:sec2}

\indent

Farhi and Gutmann [\ref{FG}] first considered quantum computation from a 
different point of view by regarding it as controlled Hamiltonian (continuous) 
time evolution of a system. This view is definitely proper, because quantum
mechanical systems naturally evolve continuously in time. We inherit their
point of view in this section. 

Let an unsorted database consist of $N$  objects $\{w_j\mid 1\le j \le N\}$, and 
let $f$ be an {\em oracle\/} (or Boolean) function such that
\begin{equation}\label{grov:eq2.1}
f(w_j) = \left\{\begin{array}{ll}
1,&j=1,2,\ldots,\ell,\\
0,&j=\ell+1, \ell+2,\ldots, N.\end{array}\right.
\end{equation}
Here the $\ell$ elements $\{w_j\mid 1\le j\le \ell\}$ are the desired objects of 
search. However, in general $\ell$ is {\em not\/} explicitly given. Note that 
the assignment of the ``desired'' objects to the first $\ell$ values of the 
index $j$ is just a convention for the purposes of theoretical discussion; from 
the point of view of the user of the algorithm, the $N$ objects are in random or 
unknown order, or perhaps better, have no meaningful ordering whatever. Consider 
now a Hilbert space $\mathcal{H}$ of dimension $N$ with an orthonormal basis 
$\mathcal{B}=\{|w_j\rangle \mid 1 \le j \le N\}$; each $|w_j\rangle$ is an
eigenstate in the 
quantum computer representing the object $w_j$ in the database. Denote $L = 
\text{span}\{|w_j\rangle\mid 1\le j\le \ell\}$. Here we have adopted the 
notation of $\text{\sl ket } |\cdot\rangle$ and $\text{\sl bra } \langle\cdot|$ 
in 
mathematical physics to denote, respectively, vectors and linear functionals in 
$\mathcal{H}$. Suppose we are given a Hamiltonian $H$ in $\mathcal{H}$ and we 
are told that $H$ has an eigenvalue $E\ne 0$ on the
entire subspace 
$L$ defined above and all the other eigenvalues are zero. The task is to find
an  eigenvector $|w_j\rangle$ in $L$ that has eigenvalue $E$. The task is
regarded  as completed when a {\em measurement\/} of the  system shows that it
is in the  state $|w_j\rangle$ for some $j\colon \ 1\le j \le \ell$.

Define a linear operator $H_L$, whose action on a basis element is given by
\begin{equation}\label{grov:eq2.2}
H_L|w_j\rangle = \frac{E}2 (|w_j\rangle -(-1)^{f(w_j)} |w_j\rangle),\qquad 
j=1,2,\ldots, N.
\end{equation}
Note that here we have only utilized the knowledge of $f$; no knowledge of the 
desired search objects $\{|w_j\rangle\mid j=1,2,\ldots,\ell\}$ is required or 
utilized since it is assumed to be hidden in the oracle (black box). 
Nevertheless, since $H_L$ is a linear operator, through {\em linear extension\/} 
we know that $H_L$ is uniquely defined, and it necessarily has the following 
unique ``explicit'' representation
\begin{equation}\label{grov:eq2.3}
H_L = E \sum^\ell_{j=1} |w_j\rangle \langle w_j|.
\end{equation}
The explicitness of $H_L$ in (\ref{grov:eq2.3}) is somewhat misleading. We need 
to emphasize that $\ell$ in (\ref{grov:eq2.3}) is {\em not explicitly\/} known 
or given since the only knowledge we have is $f$ in (\ref{grov:eq2.1}). For the 
implementation of the algorithms here as well as in the next section, this does 
not constitute any problem, however, except that in most applications the 
determination of and the information about $\ell$ are important. This becomes a 
separate class of problems calling {\em counting\/} that is studied in 
[\ref{BHT}], which we hope to expound further in a sequel. As in [\ref{FG}],
we now add to $H_L$ the ``driving Hamiltonian''
\begin{equation}\label{grov:eq2.4} H_D = E|s\rangle \langle s|
\end{equation}
for some (yet  arbitrary) unit vector $|s\rangle \in \mathcal{H}$, $|s\rangle 
\notin L$. This gives the overall Hamiltonian as
\begin{equation}\label{grov:eq2.5}
H = H_L + H_D.
\end{equation}
Our quantum computer is governed by the Schr\"odinger equation
\begin{equation}\label{grov:eq2.6}
\left\{\begin{array}{ll}
i \frac{d}{dt} |\psi(t)\rangle = H|\psi(t)\rangle,&t>0,\\
|\psi(0)\rangle = |s\rangle,\end{array}\right.
\end{equation}
as a continuous-time, controlled Hamiltonian system. Since 
(\ref{grov:eq2.6})$_1$ is autonomous, the state of the system at time $t$ is 
given by 
\begin{equation}\label{grov:eq2.7}
|\psi(t)\rangle = e^{-iHt} |s\rangle,\qquad t\ge 0,
\end{equation}
where $e^{-iHt}$ is the exponential $N\times N$ 
(time evolution) matrix.

Define an augmented space $\wt L$ from $L$ and $|s\rangle$:
\begin{equation}\label{grov:eq2.8}
\wt L = \text{span}(L\cup \{|s\rangle\}).
\end{equation}
Let $\wt L^\bot$ be the orthogonal complement of $\wt L$ satisfying $\wt L 
\oplus \wt L^\bot = \mathcal{H}$, where $\oplus$ denotes the orthogonal direct 
sum. With respect to this orthogonal decomposition, we now have our first 
reduction of dimensionality below.

\begin{prop}\label{grov:prop2.1}
Fix $|s\rangle\in \mathcal{H}$, $|s\rangle\notin L$ in (\ref{grov:eq2.6}). Let 
$H, \wt L$ and $\wt L^\bot$ be defined as above. For any $|w\rangle \in 
\mathcal{H}$, write $|w\rangle = |v\rangle + |u\rangle \in \wt L \oplus \wt 
L^\bot$ according to the orthogonal direct sum. Then $H|w\rangle = H|v\rangle$ 
and $H|u\rangle = 0$. Consequently, the Hamiltonian $H$ has an associated 
blockwise decomposition
\begin{equation}\label{grov:eq2.9}
H = \left[\begin{matrix}
H_{\wt L}&\vdots&0_{12}\\
\hdotsfor{3}\\
0_{21}&\vdots&0_{22}\end{matrix}\right]
\end{equation}
where $H_{\wt L}$ is an invertible $(\ell+1) \times (\ell+1)$ matrix defined on 
$\wt L$ such that $H_{\wt L}|v\rangle = H|v\rangle$ for all $|v\rangle \in \wt 
L$, and $0_{12}, 0_{21}$ and $0_{22}$ are, respectively, $(\ell+1) \times 
(N-\ell-1)$, $(N-\ell-1)\times (\ell +1)$ and $(N-\ell-1) \times (N-\ell-1)$ 
zero matrices.
\end{prop}

\begin{proof}
Straightforward verification.
\end{proof}

\begin{cor}\label{grov:cor2.2}
Fix $|s\rangle \in \mathcal{H}, |s\rangle\notin L$ in (\ref{grov:eq2.6}). Let 
$H,\wt L$ an $\wt L^\bot$ be defined as above. Then the state of the solution of 
(\ref{grov:eq2.6}) at time $t, |\psi(t)\rangle$, has zero component in $\wt 
L^\bot$ for all $t>0$.
\end{cor}

\begin{proof} 
The action of the evolution dynamics $e^{-iHt}$ on the invariant subspace $\wt 
L^\bot$ is, by (\ref{grov:eq2.9}), 
$e^{-i0_{22}t} = e^0 = \pmb{I}_{\wt 
L^\bot}$, the identity operator on $\wt L^\bot$. Since the component of 
$|s\rangle$ in $\wt L^\bot$ is the zero vector, the action of 
$\pmb{I}_{N-\ell-1}$  
(the $(N-\ell-1)\times(N-\ell-1)$ identity matrix)
on it remains zero for all $t>0$.
\end{proof}

By the properties obtained above, we need to fix our attention only on $H_{\wt 
L}$ defined on $\wt L$. By abuse of notation, we still write $H$ instead of 
$H_{\wt L}$ on $\wt L$.

\begin{prop}[Matrix representation of $\pmb{H}$ on $\pmb{\wt 
L}$]\label{grov:prop2.3}
Under the same assumptions as in Prop.~\ref{grov:prop2.1}, 
define $x_i = \langle 
s|w_i\rangle$ for $i=1,2,\ldots, \ell$. Let 
\begin{equation}\label{grov:eq2.10}
|r\rangle = \frac1{C_r} \left(|s\rangle - \sum^\ell_{i=1}
x_i|w_i\rangle\right),  \qquad C_r\equiv \sqrt{1 -\sum\limits^\ell_{i=1}
|x_i|^2}. \end{equation}
Then $\{|w_1\rangle,\ldots, |w_\ell\rangle, |r\rangle\}$ forms an
orthonormal basis of 
$\wt L$ with respect to which $H$ admits the matrix representation
\begin{equation}\label{grov:eq2.11}
H = E[H_{ij}]_{(\ell+1)\times (\ell+1)};\quad H_{ij} = \left\{\begin{array}{ll}
x_j\bar x_i,&1 \le i,j\le \ell, i\ne j,\\
1+|x_j|^2,&1\le i,j\le \ell, i=j,\\
(\delta_{j,\ell+1}x_j + \delta_{i,\ell+1} \bar x_i) C_r,&i=\ell+1 \text{ or } 
j=\ell+1, i\ne j,\\
C^2_r,&i=j=\ell+1.\end{array}\right.
\end{equation}
\end{prop}

\begin{proof}
Solve (\ref{grov:eq2.10}) for $|s\rangle$:
\begin{equation}\label{grov:eq2.12}
|s\rangle = \sum^\ell_{i=1} x_i|w_i\rangle + C_r|r\rangle. 
\end{equation}
Substituting (\ref{grov:eq2.12}) into (\ref{grov:eq2.3}) and
(\ref{grov:eq2.4}), we obtain (\ref{grov:eq2.11}) in bra-ket form.
  \end{proof}

The exponential matrix function $e^{-iHt}$ on $\wt L$ based upon the 
representation can be obtained, e.g., by
\begin{equation}\label{grov:eq2.13}
e^{-iHt} = \sum^\infty_{k=0} \frac{t^k}{k!} (-iH)^k
\end{equation}
or
\begin{equation}\label{grov:eq2.14}
e^{-iHt} = \frac1{2\pi i} \oint\limits_{\mathcal{C}} (\zeta \pmb{I}_{\ell+1} + 
iH)^{-1} e^{\zeta t} d\zeta
\end{equation}
where $\mathcal{C}$ is a simple closed curve in $\mathbb{C}$ enclosing all the 
values $\zeta$ such that the $(\ell+1)\times (\ell+1)$ matrix $\zeta
\pmb{I}_{\ell+1} +iH$ 
is not invertible. However, since $\ell$ can be arbitrarily large, it is a 
highly nontrivial task (if possible at all) to calculate $e^{-iHt}$ explicitly 
based on (\ref{grov:eq2.13}), (\ref{grov:eq2.14}) or any other known techniques.

It turns out that the above difficulty can be bypassed if $|s\rangle$ is chosen 
in an ingenious way that can effect another reduction of dimensionality. 
(Nevertheless, we again call attention to the fact that any choice of 
$|s\rangle$ must be independent of any knowledge of $|w_i\rangle$, for 
$i=1,2,\ldots, \ell$.) The choice by Grover [\ref{grover1}, \ref{grover2}] is 
\begin{equation}\label{grov:eq2.15}
|s\rangle = \frac1{\sqrt N} \sum^N_{j=1} |w_j\rangle,
\end{equation}
implementable and obtainable on the QC through an application of the 
Walsh-Hadamard transformation on all the qubits; $|s\rangle$ is a superposition 
state with the same amplitude in all eigenstates. With this choice of 
$|s\rangle$, we now have
\begin{equation}\label{grov:eq2.16}
x_i = x \equiv 1/\sqrt N, i=1,2,\ldots,\ell;\quad C^2_r = 1-(\ell/N),\quad 
|r\rangle = \frac1{\sqrt{1-(\ell/N)}} \left(|s\rangle - \frac1{\sqrt N} 
\sum^\ell_{i=1} |w_i\rangle\right)
\end{equation}
in (\ref{grov:eq2.11}) and (\ref{grov:eq2.10}).

\begin{thm}[A two-dimensional invariant subspace for the Hamiltonian 
$\pmb{H}$]\label{grov:thm2.4}
Let $|s\rangle$ be given as in (\ref{grov:eq2.15}). Denote
\begin{equation}\label{grov:eq2.16a}
\mathcal{V} = \{|v\rangle\in \wt L\ \Big| \ |v\rangle = a \sum^\ell_{i=1} 
|w_i\rangle + b|r\rangle;\quad a,b\in \mathbb{C}\}.
\end{equation}
Then $\mathcal{V}$ is a invariant two-dimensional subspace of $H$ in 
$\wt L$ such that
\begin{itemize}
\item[(1)] $r,s,\in \mathcal{V}$;
\item[(2)] $H(\mathcal{V}) = \mathcal{V}$.
\end{itemize}
\end{thm}

\begin{proof}
It is obvious that (1) holds. To see (2), we have
\begin{align}
H|v\rangle &\equiv H \left[a \sum^\ell_{i=1} |w_i\rangle +b|r\rangle\right] 
\nonumber\\
&= a\sum^\ell_{i=1} H|w_i\rangle + bH|r\rangle\nonumber\\
&= E\Bigg\{a \sum^\ell_{i=1} \left[(1+x^2) |w_i\rangle + x^2 \sum^\ell_{\sst 
j=1\atop
\sst j\ne i} |w_j\rangle + x\sqrt{1-\ell x^2} 
|r\rangle\right]\nonumber\\
&\quad  + b \left[x\sqrt{1-\ell x^2} \sum^\ell_{j=1} |w_j\rangle + (1-\ell 
x^2) |r\rangle\right]\Bigg\}\nonumber\\
&= E \left\{a \sum^\ell_{i=1} \left[|w_i\rangle + x^2 \sum^\ell_{j=1} 
|w_j\rangle\right]\right.\nonumber\\
&\quad \left. + bx\sqrt{1-\ell x^2} \sum^\ell_{j=1} |w_j\rangle + [a\ell 
x\sqrt{1-\ell x^2} + b(1-\ell x^2)]|r\rangle\right\}\nonumber\\
\label{grov:eq2.17}
&= E\left\{(a+a\ell x^2 + bx\sqrt{1-\ell x^2} \sum^\ell_{i=1} |w_i\rangle + 
[a\ell x\sqrt{1-\ell x^2} + b(1-\ell x^2)]|r\rangle\right\}\in \mathcal{V}.
\end{align}
\end{proof}

\begin{cor}\label{grov:cor2.4}
Define
\begin{equation}\label{grov:eq2.17a}
|\wt w\rangle = \frac1{\sqrt\ell} \sum^\ell_{i=1} |w_i\rangle.
\end{equation}
Then $\{|\wt w\rangle, |r\rangle\}$ forms an orthonormal basis for $\mathcal{V}$ 
such that with respect to this basis, $H$ admits the matrix representation
\begin{equation}\label{grov:eq2.18}
H = E \left[\begin{matrix}
1+\frac\ell{N}&\frac{\sqrt{\ell(N-\ell)}}{N}\\
\frac{\sqrt{\ell(N-\ell)}}{N}&1-\frac\ell{N}\end{matrix}\right]
\end{equation}
with
\begin{equation}\label{grov:eq2.19}
e^{-iHt} = e^{-iEt} \left[\begin{matrix}
\cos (Eyt) - iy \sin(Eyt)& -\sqrt{1-y^2} i\sin (Eyt)\\
-\sqrt{1-y^2} i\sin (Eyt)&\cos(Eyt) + iy \sin(Eyt) 
\end{matrix}\right], y\equiv \sqrt{\ell/N}.
\end{equation}
\end{cor}

\begin{proof}
The representation (\ref{grov:eq2.18}) follows easily from (\ref{grov:eq2.17}).

To calculate $e^{-iHt}$, write
$$H = E\left[\begin{matrix} 1+y^2&y\sqrt{1-y^2}\\ 
y\sqrt{1-y^2}&1-y^2\end{matrix}\right],\qquad y\equiv \sqrt{\ell/N},$$
analogous to [\ref{FG}, (8)], and apply (\ref{grov:eq2.14}) to obtain 
(\ref{grov:eq2.19}); or apply (\ref{grov:eq2.13}) and the properties of the
$SU(2)$ generators commonly used in quantum mechanics [\ref{Me}, (XIII.84),
p.~546].
 \end{proof}

Since $\mathcal{V}$ is invariant under $H$ and $H$ is a self-adjoint matrix, the 
orthogonal complement $\mathcal{V}^\bot$ of $\mathcal{V}$ is also an invariant 
subspace of $H$. The precise action of $H$ or $e^{iHt}$ on $\mathcal{V}^\bot$ 
does not seem to be describable in simple terms. However, this knowledge is not 
needed as we now have the explicit form of the solution available below.

\begin{cor}\label{grov:cor2.5}
The solution $\psi(t)$ of (\ref{grov:eq2.6}) given (\ref{grov:eq2.15}) is
\begin{equation}\label{grov:eq2.20}
\psi(t) = e^{-iEt} \{[y\cos(Eyt) - i \sin(Eyt)] |\wt w\rangle + \sqrt{1-y^2} 
\cos(Eyt)|r\rangle\},\quad t>0,
\end{equation}
where $y = \sqrt{\ell/N}$.
\end{cor}

\begin{proof}
Use (\ref{grov:eq2.19}) and (\ref{grov:eq2.16}).
\end{proof}

Note that (\ref{grov:eq2.20}) has the same form as [\ref{FG}, (10)]. Since $|\wt 
w\rangle$ and $|r\rangle$ are unit, mutually orthogonal vectors, the probability 
of reaching the state $|\wt w\rangle$ at time $t$ is
\begin{equation}\label{grov:eq2.21}
P(t) = \sin^2(Eyt) + y^2\cos^2(Eyt),\qquad y=\sqrt{\ell/N}.
\end{equation}
At time $T\equiv \pi/(2Ey)$, the probability is 1. By (\ref{grov:eq2.17a}), if 
we make a measurement of $|\psi(t)\rangle$ at $t=T$, we will obtain any one of 
the eigenstates $|w_i\rangle$, $i=1,2,\ldots,\ell$, with equal probability 
$1/\ell$. Therefore the task of search is completed ({\em with probability 1}), 
requiring a total time duration
\begin{equation}\label{grov:eq2.22}
T = \pi/(2Ey) = \frac\pi{2E} \sqrt{\frac{N}\ell}.
\end{equation}
Formula (\ref{grov:eq2.21}) manifests the notion of {\em amplitude 
amplication\/} because the amplitude of $|\psi(t)\rangle$ along $|r\rangle$, 
$\sqrt{1-y^2} \cos(Eyt)$, is steadily decreasing in magnitude as a function of 
$t\in [0,T]$, implying that $P(t)$ in (\ref{grov:eq2.21}) is increasing for 
$t\in [0,T]$. 

Next, let us address the optimality of the search algorithm as given above. We 
assume that $N/\ell$ is a large number. We show that the above generalized 
Grover-Farhi-Gutmann algorithm for multiobject search is time-optimal within the 
order $\mathcal{O}(\sqrt{N/\ell})$. In contrast with the classical random
search which requires in average $(N+1)/(\ell+1)$ searches (see (\ref{eqA.6})
in the Appendix), again we see that there is a quadratic speedup.

The idea of proof  follows by combining 
those in [\ref{BBHT}] and [\ref{FG}].
Let the Hamiltonian $H_L$ be given as in (\ref{grov:eq2.3}). We wish to add a 
somewhat arbitrary (generally, time-dependent) driving Hamiltonian $H_D(t)$ to 
$H_L$ so that the terminal state (at time $\wt T$) is in $L$ (which, after a 
measurement, becomes one of the eigenstates $|w_1\rangle, |w_2\rangle,\ldots, 
|w_\ell\rangle$). Our objective is to find a lower bound on $\wt T$.
Of course,  $H_D(t)$ and $\wt T$ must be independent of $L$,
since they are part of the algorithm prescribed for determining~$L$.

Let $|w_I\rangle \in \mathcal{H}$ be an arbitrary initial state such that 
$\langle w_I| w_I\rangle = 1$. Let $\psi_L(t)$ denote the solution of the 
Schr\"odinger equation
\begin{equation}\label{grov:eq2.23}
\left\{\begin{array}{l}
i \ds\frac{d}{dt} |\psi_L(t)\rangle = (H_L + H_D(t)) |\psi_L(t)\rangle,\qquad 
0<t\le \wt T,\\
|\psi_L(0)\rangle = |w_I\rangle,\qquad \text{(initial condition)}\\
|\psi_L(\wt T)\rangle = \ds\sum\limits^\ell_{i=1} \alpha_i|w_i\rangle\in 
L,\qquad 
\text{(terminal condition).}
\end{array}\right.
\end{equation}
Note that $\alpha_i \in \mathbb{C}$ for $i=1,2,\ldots, \ell$ and
\begin{equation}\label{grov:eq2.24}
\sum^\ell_{i=1} |\alpha_i|^2 =1
\end{equation}
because the evolution process is unitary at any $t\in (0,\wt T]$. On the other 
hand, let $|\psi(t)\rangle$ evolve with $H_D(t)$:
\begin{equation}\label{grov:eq2.25}
\left\{\begin{array}{ll}
i\ds\frac{d}{dt} |\psi(t)\rangle = H_D(t) |\psi(t)\rangle,&0< t \le \wt T,\\
|\psi(0)\rangle = |w_I\rangle.\end{array}\right.
\end{equation}

\begin{lem}\label{grov:lem2.6}
Assume that $\wt N \equiv N/\ell$ is an integer. Let the orthonormal basis
$\mathcal{B} = 
\{|w_i\rangle\mid 1\le i \le N\}$ be grouped into $\wt N$ disjoint subsets
\begin{equation}\label{grov:eq2.26}
\mathcal{B} = \dot{\bigcup\nolimits}^{\wt N}_{k=1} B_k
\end{equation}
where each $B_k \equiv \{|w_{i,k}\rangle \mid i=1,2,\ldots,\ell\}$ contains 
exactly $\ell$ orthonormal basis elements. Then we have
\begin{align}
2\wt N -2\sqrt{\wt N} &\le \sum^{\wt N}_{k=1} \langle\psi_{L_k} (\wt T) - 
\psi(\wt T)| \psi_{L_k}(\wt T) - \psi(\wt T)\rangle\nonumber\\
\label{grov:eq2.27}
&\le 2E\sqrt{\wt N} \ \wt T.
\end{align}
where $|\psi_{L_k}(t)\rangle$ is the solution $|\psi_L(t)\rangle$ of 
(\ref{grov:eq2.23}) with $L =L_k = \text{span } B_k$
in both the differential equation and the terminal condition.
 Consequently,
\begin{equation}\label{grov:eq2.28}
\wt T \ge (1-\vp_{\wt N}) \frac{\wt N^{1/2}}E = \frac{1-\vp_{\wt N}}E 
\sqrt{\frac{N}\ell},
\end{equation}
where $\vp_{\wt N} = \wt N^{-1/2}\to 0$ as $\wt N\to \infty$.
\end{lem}

\begin{proof}
As in [\ref{FG}, (1.8)], we have
\begin{equation}\label{grov:eq2.29}
\langle\psi_{L_k}(\wt T) - \psi(\wt T)|\psi_{L_k}(\wt T) - \psi(\wt T)\rangle = 
2 -\sum^\ell_{i=1} [\langle \alpha_{i,k} w_{i,k}| \psi(\wt T)\rangle + \langle 
\psi(\wt T)| \alpha_{i,k} w_{i,k}\rangle],
\end{equation}
and, therefore
$$\sum^{\wt N}_{k=1} \langle \psi_{L_k}(\wt T) - \psi(\wt T)|\psi_{L_k}(\wt T) - 
\psi(\wt T)\rangle = 2\wt N - \sum^{\wt N}_{k=1} \left[\Big\langle 
\sum^\ell_{i=1} 
\alpha_{i,k}w_{i,k} \Big| \psi(\wt T)\Big\rangle 
+ \Big\langle\psi(\wt T) \Big| 
\sum^\ell_{i=1} \alpha_{i,k} w_{i,k}\Big\rangle\right].$$
Let $\mathbb{P}_{L_k}\colon \ \mathcal{H}\to L_k$ be the orthogonal projection 
of $\mathcal{H}$ onto $L_k$. Then
\begin{equation}\label{grov:eq2.30}
\sum^{\wt N}_{k=1} \bigg|\Big\langle \sum^\ell_{i=1} \alpha_{i,k} 
w_{i,k}\Big| \psi(\wt T)\Big\rangle\bigg|^2 \le \sum^{\wt N}_{k=1} 
|\mathbb{P}_{L_k} 
|\psi(\wt T)\rangle\|^2\le 1,
\end{equation}
from which, we apply the Cauchy-Schwarz inequality and obtain
\begin{equation}\label{grov:eq2.31}
\sum^{\wt N}_{k=1} \bigg|\Big\langle \sum^\ell_{i=1} \alpha_{i,k} w_{i,k} 
\Big|\psi(\wt T)\Big\rangle + \Big\langle \psi(\wt T)\Big| \sum^\ell_{i=1} 
\alpha_{i,k} 
w_{i,k}\Big\rangle\bigg| \le 2\wt N^{1/2}.
\end{equation}
Combining (\ref{grov:eq2.29}) and (\ref{grov:eq2.31}), we have established the 
left half of the inequality (\ref{grov:eq2.27}).

Next, mimicking [\ref{FG}, (19)--(21)], we have
\begin{align*}
\frac{d}{dt} [\langle \psi_{L_k}(t) - \psi(t)| \psi_{L_k}(t) - \psi(t)\rangle] 
&= 2 \text{ Im}\langle \psi_{L_k}(t)|H_{L_k}|\psi(t)\rangle\\
&\le 2|\langle \psi_{L_k}(t)|H_{L_k}|\psi(t)\rangle|\\
&\le 2\|H_{L_k}|\psi(t)\rangle \| = 2E\left[ \sum^\ell_{i=1} |\langle w_{i,k} 
|\psi(t)\rangle|^2\right]^{1/2} \\
&= 2E|\mathbb{P}_{L_k}|\psi(t)\rangle|,\\
\frac{d}{dt} \sum^{\wt N}_{k=1} [\langle \psi_{L_k} (t) - \psi(t)| \psi_{L_k}(t) 
- \psi(t)\rangle] &\le 2E \sum^{\wt N}_{k=1} |\mathbb{P}_{L_k}|\psi(t) \rangle|,
\end{align*}
and from $\sum\limits^{\wt N}_{k=1} |\mathbb{P}_{L_k} |\psi(t)\rangle|^2 = 1$, 
by an application of the Cauchy-Schwarz inequality again, we obtain
\begin{equation}\label{grov:eq2.32}
\frac{d}{dt} \sum^{\wt N}_{k=1} [\langle \psi_{L_k}(t) - \psi(t)|\psi_{L_k} (t) 
- \psi(t)\rangle] \le 2E \wt N^{1/2}.
\end{equation}
Integrating (\ref{grov:eq2.32}) from 0 to $\wt T$, noting that $|\psi_{L_k}(0) - 
\psi(0)\rangle=0$, we have verified the right half of inequality 
(\ref{grov:eq2.27}).
\end{proof}

In general, $N/\ell$ is not necessarily an integer. Therefore, the disjoint 
union (\ref{grov:eq2.26}) is not always possible. Define $\wt N = [N/\ell]$, 
namely, the integral part of the rational number $N/\ell$. Then
$$\frac{N}\ell = \wt N +\delta,\quad \text{where}\quad 0\le \delta <1.$$
Then we can rewrite (\ref{grov:eq2.26}) as
\begin{equation}\label{grov:eq2.33}
\mathcal{B} = \dot{\bigcup\nolimits}^{\wt N}_{k=1} B_k \cup R,
\end{equation}
where $\dot{\bigcup}^{\wt N}_{k=1} B_k$ is an arbitrary collection of disjoint 
sets of $\ell$ orthonormal basis elements containing a total of $\wt N\ell$ of 
them, and $R$ is the remaining set of orthonormal basis elements with 
cardinality $\ell\delta (<\ell)$. The proof of Lemma 2.7 extends to this case 
except for some 
tedious details of bookkeeping concerning the short set $R$, 
which we omit.
 We therefore have arrived at 
the following order of time-optimality for the continuous-time generalized 
Grover algorithm for multiobject search.

\begin{thm}\label{grov:thm2.7}
Assume that $N/\ell$ is large. Then it requires at least
$$\wt T = \frac{1-\vp_{N/\ell}}{E} \sqrt{\frac{N}\ell},\quad \vp_{N/\ell} 
>0,\quad \vp_{N/\ell} = \mathcal{O}((N/\ell)^{-1/2}),$$
time duration in average
for the driven general Hamiltonian system (\ref{grov:eq2.25}) to 
reach the subspace $L$.$\hfill\square$
\end{thm}

\section{Discrete Time Case: \ Straightforward Generalization of Grover's 
Algorithm to Multiobject Search}\label{grov:sec3}

\setcounter{equation}{0}

\indent

In this section we generalize Grover's search algorithm in its original 
form [\ref{grover1}, 
\ref{grover2}] to the situation where the number of objects 
satisfying the search criterion is greater than 1. We are considering the
discrete time case here, which may be regarded as a discrete-time sampled
system of the continuous-time case treated in the preceding section. Unlike the
continuous-time case, there have been a relatively rich literature studying
the generalization of Grover's discrete-time algorithm to multiobject search
(see [\ref{Bih}, \ref{Bir}, \ref{BBHT}, \ref{BHT}] and the references therein).
Our presentation below gives more formalized arguments than those earlier
contributions, provides a clearer Hilbert space framework, and settles a
relevant question contained in [\ref{BBHT}].

Let the database $\{w_i\mid i=1,2,\ldots, N\}$, orthonormal eigenstates 
$\{|w_i\rangle \mid i=1,2,\ldots, N\}$ and the oracle function  $f$ be the same 
as given at the beginning of \S 2. The definitions of $\mathcal{H}, L$ and 
$\ell$ remain the same.

Define a linear operation in terms of the oracle function $f$ as follows:
\begin{equation}\label{grov:eq3.1}
I_L|w_j\rangle = (-1)^{f(w_j)} |w_j\rangle,\qquad j=1,2,\ldots, N.
\end{equation}
Then since $I_L$ is linear, the extension of $I_L$ to the entire space 
$\mathcal{H}$ is unique, with an ``explicit'' representation
\begin{equation}\label{grov:eq3.2}
I_L = \pmb{I} - 2 \sum^\ell_{j=1} |w_j\rangle \langle w_j|,
\end{equation}
where $\pmb{I}$ is the identity operator on $\mathcal{H}$. $I_L$ is the operator 
of {\em rotation (by $\pi$) of the phase\/} of the subspace $L$. Note again that 
the explicitness of (\ref{grov:eq3.2}) is misleading because explicit knowledge 
of $\{|w_j\rangle\mid 1\le j \le \ell\}$ and $\ell$ in (\ref{grov:eq3.2}) is not 
available. Nevertheless, (\ref{grov:eq3.2}) is a well-defined ({\em unitary\/}) 
operator on $\mathcal{H}$ because of (\ref{grov:eq3.1}). (Unitarity  
is a requirement for all operations in a QC.)

We now define $|s\rangle$ as in (\ref{grov:eq2.15}). Then
\begin{equation}\label{grov:eq3.3}
|s\rangle = \frac1{\sqrt N} \sum^N_{i=1} |w_i\rangle = \frac1{\sqrt N} 
\sum^\ell_{i=1} |w_i\rangle + \sqrt{\frac{N-\ell}N} |r\rangle; \text{ see 
(\ref{grov:eq2.16}) for } |r\rangle.
\end{equation}
Now, define another operator, {\em the inversion about average operation}, just 
as in Grover [\ref{grover1}, \ref{grover2}]:
\begin{equation}\label{grov:eq3.4}
I_s = \pmb{I} -2|s\rangle \langle s|.
\end{equation}
Note that $I_s$ in (\ref{grov:eq3.4}) is unitary and hence quantum mechanically 
admissible. $I_s$ is  {\em explicitly known}, constructible with the
so-called Walsh-Hadamard transformation.

\begin{lem}\label{grov:lem3.1}
Let $\wt L$ be defined as in (\ref{grov:eq2.8}). Then $\{|w_i\rangle, 
|r\rangle\mid 
i=1,2,\ldots, \ell\}$ forms an orthonormal basis of $\wt L$. The orthogonal 
direct sum $\mathcal{H} = \wt L \oplus \wt L^\bot$ is an orthogonal invariant 
decomposition for both operators $I_{\wt L}$ and $I_s$. Furthermore,
\begin{itemize}
\item[(i)] The restriction of $I_s$ on $\wt L$ admits a unitary matrix 
representation with respect to the orthonormal basis $\{|w_1\rangle, 
|w_2\rangle,\ldots, |w_\ell\rangle, |r\rangle\}$:
\begin{align}
A &= [a_{ij}]_{(\ell+1)\times (\ell+1)},\nonumber\\
\label{grov:eq3.5}
a_{ij} &= \left\{\begin{array}{ll}
\delta_{ij} - \ds\frac2N,&1\le i,j\le \ell,\\
-\ds\frac{2\sqrt{N-\ell}}N (\delta_{i,\ell+1} + \delta_{j,\ell+1}),& i=\ell+1 
\text{ or } j=\ell+1, i\ne j,\\
\ds\frac{2\ell}N -1,&i=j=\ell+1.\end{array}\right.
\end{align}
\item[(ii)] The restriction of $I_s$ of $\wt L^\bot$ is $\mathbb{P}_{\wt 
L^\bot}$, 
the orthogonal projection operator on $\wt L^\bot$. Consequently, $I_s|_{\wt 
L^\bot} = \pmb{I}_{\wt L^\bot}$, where $\pmb{I}_{\wt L^\bot}$ is the identity 
operator on $\wt L^\bot$.
\end{itemize}
\end{lem}

\begin{proof}
We have, from (\ref{grov:eq3.3}) and (\ref{grov:eq3.4}),
\begin{align}
I_s &= \pmb{I} -2\left[\frac1{\sqrt N} \sum^\ell_{i=1} |w_i\rangle + 
\sqrt{\frac{N-\ell}N} |r\rangle\right] \left[\frac1{\sqrt N} \sum^\ell_{j=1} 
\langle w_j| + \sqrt{\frac{N-\ell}N} \langle r|\right]\nonumber\\
&= \left[\sum^\ell_{i=1} |w_i\rangle \langle w_i\rangle + |r\rangle \langle r| + 
\mathbb{P}_{\wt L^\bot}\right] - \left\{\frac2N \sum^\ell_{i=1} \sum^\ell_{j=1} 
|w_i\rangle \langle w_j|\right.\nonumber\\
&\quad \left. + \frac{2\sqrt{N-\ell}}N \left[\sum^\ell_{i=1} (|w_i\rangle 
\langle r| + |r\rangle \langle w_i|)\right] + 2\left(\frac{N-\ell}N\right) 
|r\rangle \langle r|\right\}\nonumber\\
&= \sum^\ell_{i=1} \sum^\ell_{j=1} \left(\delta_{ij} - \frac2N\right) 
|w_i\rangle \langle w_j| - \frac{2\sqrt{N-\ell}}N \left[\sum^\ell_{i=1} 
(|w_i\rangle \langle r| + |r\rangle \langle w_i|)\right]\nonumber\\
\label{grov:eq3.6}
&\quad + \left(\frac{2\ell}N-1\right) |r\rangle \langle r| + \mathbb{P}_{\wt 
L^\bot}.
\end{align}
The conclusion follows.
\end{proof}

The generalized ``Grover's search engine'' for multiobject search is now defined 
as
\begin{equation}\label{grov:eq3.7}
U  = -I_sI_L.
\end{equation}

\begin{lem}\label{grov:lem3.2}
The orthogonal direct sum $\mathcal{H} = \wt L \oplus \wt L^\bot$ is an 
invariant decomposition for the unitary operator $U$, such that the
following 
holds:
\begin{itemize}
\item[(1)] With respect to the orthonormal basis $\{|w_1\rangle,\ldots, 
|w_\ell\rangle, |r\rangle\}$ of $\wt L,U$ admits a unitary matrix representation
\begin{align}
U|_{\wt L} &= [u_{ij}]_{(\ell+1)\times (\ell+1)},\nonumber\\
\label{grov:eq3.8}
u_{ij} &= \left\{\begin{array}{ll}
\delta_{ij} - \ds\frac2N,&1\le i,j\le \ell,\\
\ds\frac{2\sqrt{N-\ell}}N (\delta_{j,\ell+1} - \delta_{i,\ell+1}),&i=\ell+1 
\text{ or } j=\ell+1, i\ne j,\\
1-\ds\frac{2\ell}N,&i=j=N+1.\end{array}\right.
\end{align}
\item[(2)] The restriction of $U$ on $\wt L^\bot$ is $-\mathbb{P}_{\wt 
L^\bot} = -\pmb{I}_{\wt L^\bot}$.
\end{itemize}
\end{lem}

\begin{proof}
Substituting (\ref{grov:eq3.2}) and (\ref{grov:eq3.6}) into (\ref{grov:eq3.7}) 
and simplifying, we obtain
\begin{align*}
U &= -I_sI_L = \cdots \text{(simplification)}\\
&= \sum^\ell_{i=1} \sum^\ell_{j=1} \left(\delta_{ij} - \frac2N\right) 
|w_i\rangle \langle w_j| + \frac{2\sqrt{N-\ell}}N \sum^\ell_{i=1} (|w_i\rangle 
\langle r| - |r\rangle \langle w_i|)\\
&\quad + \left(1 - \frac{2\ell}N\right) |r\rangle \langle r| - \mathbb{P}_{\wt 
L^\bot}.
\end{align*}
The proof follows.
\end{proof}

Lemmas \ref{grov:lem3.1} and \ref{grov:lem3.2} above effect a reduction of the 
problem to an invariant subspace $\wt L$, just as Prop.~\ref{grov:prop2.1} did. 
However, $\wt L$ is an $(\ell+1)$-dimensional subspace where $\ell$ may also be 
large. Another reduction of dimensionality is needed to further simplify the 
operator $U$.

\begin{prop}\label{grov:prop3.3}
Define $\mathcal{V}$ as in (\ref{grov:eq2.16a}). Then $\mathcal{V}$ is an 
invariant two-dimensional subspace of $U$ such that
\begin{itemize}
\item[(1)] $r,s\in \mathcal{V}$;
\item[(2)] $U(\mathcal{V}) = \mathcal{V}$.
\end{itemize}
\end{prop}

\begin{proof}
Straightforward verification.
\end{proof}

Let $|\wt w\rangle$ be defined as in (\ref{grov:eq2.17a}). Then as in \S 2, 
$\{|\wt w\rangle, |r\rangle\}$ forms an orthonormal basis of $\mathcal{V}$. We 
have the second reduction, to dimensionality 2.

\begin{thm}\label{grov:thm3.4} 
With respect to the orthonormal basis $\{|\wt w\rangle, |r\rangle\}$ in the 
invariant subspace $\mathcal{V}, U$ admits the unitary matrix representation
\begin{equation}\label{grov:eq3.9}
U = \left[\begin{matrix}
\frac{N-2\ell}N& \frac{2\sqrt{\ell (N-\ell)}}N\\ 
-\frac{2\sqrt{\ell(N-\ell)}}N& \frac{N-2\ell}N\end{matrix}\right] = 
\left[\begin{matrix} \cos \theta & \sin \theta\\ -\sin \theta& \cos\theta 
\end{matrix}\right], \theta\equiv \sin ^{-1} \left(\frac{2\sqrt{\ell(N-\ell)}}N 
\right).
\end{equation}
\end{thm}

\begin{proof}
Use the matrix representation (\ref{grov:eq3.8}) and (\ref{grov:eq2.17a}).
\end{proof}

Since $|s\rangle\in \mathcal{V}$, we can calculate $U^m|s\rangle$ efficiently 
using (\ref{grov:eq3.9}):
\begin{align}
U^m|s\rangle &= U^m \left(\frac1{\sqrt N} \sum^\ell_{i=1} |w_i\rangle + 
\sqrt{\frac{N-\ell}N} |r\rangle\right)\qquad \text{(by (\ref{grov:eq3.3}))} 
\nonumber\\
&= U^m \left(\frac\ell{\sqrt N} |\wt w\rangle + \sqrt{\frac{N-\ell}N} 
|r\rangle\right)\nonumber\\
&= \left[\begin{matrix} \cos \theta& -\sin \theta\\ \sin \theta&\cos 
\theta\end{matrix}\right]^m \left[\begin{matrix} \frac\ell{\sqrt N}\\ 
\sqrt{\frac{N-\ell}N}\end{matrix}\right]\nonumber\\
\label{grov:eq3.10}
&= \left[\begin{matrix} \cos (m\theta+\alpha)\\ \sin(m\theta +\alpha)] 
\end{matrix}\right],\quad \left(\alpha \equiv \cos^{-1} \frac\ell{\sqrt 
N}\right)\\
&= \cos(m \theta+\alpha)\cdot |\wt w\rangle + \sin(m \theta + \alpha)\cdot 
|r\rangle.\nonumber
\end{align}
Thus, the probability of reaching the state $|\wt w\rangle$ after $m$ iterations 
is
\begin{equation}\label{grov:eq3.11}
P_m = \cos^2(m\theta +\alpha);
\end{equation}
cf.\ (\ref{grov:eq2.21}) in the continuous-time case. If $\ell \ll N$, then 
$\alpha$ is close to $\pi/2$ and, therefore, (\ref{grov:eq3.11}) is an 
increasing function of $m$ initially. This again manifests the notion of 
amplitude amplification. This probability $P_m$ is maximized if $m\theta +\alpha 
= \pi$, implying
$$m = \left[\frac{\pi-\alpha}\theta\right] = \text{the integral part of } 
\frac{\pi-\alpha}\theta.$$
When $\ell/N$ is small, we have
\begin{align*}
\theta &= \sin^{-1} \left(\frac{2\sqrt{\ell(N-\ell)}}N\right)\\
&= \sin^{-1} \left(2 \sqrt{\frac\ell{N}} \left[1-\frac12 \frac\ell{N} - \frac18 
\left(\frac\ell{N}\right)^2 \pm\cdots\right]\right)\\
&= 2 \sqrt{\frac\ell{N}} + \mathcal{O}((\ell/N)^{3/2}),\\
\alpha &= \cos^{-1} \frac\ell{\sqrt N} = \frac\pi2 - \left[\frac\ell{\sqrt N} + 
\mathcal{O} ((\ell/N^{1/2})^3)\right].
\end{align*}
Therefore
\begin{align}
m &\approx \frac{\pi- \left\{\frac\pi2 - \left[\frac\ell{\sqrt N} + 
\mathcal{O}((\ell/N^{1/2})^3)\right]\right\}}{2\sqrt{\frac\ell{N}} + \mathcal{O} 
((\ell/N)^{3/2})}\nonumber\\
\label{grov:eq3.12}
&= \frac\pi4 \sqrt{\frac{N}\ell} \left[1 + \mathcal{O}\left(\frac\ell{N} 
\right)\right].
\end{align}

\begin{cor}\label{grov:cor3.5}
The generalized Grover's algorithm for  multiobject search with operator 
$U$ 
given by (\ref{grov:eq3.7}) has success probability $P_m = \cos^2(m\theta 
+\alpha)$ of reaching the state $|\wt w\rangle \in L$ after $m$ 
iterations. For $\ell/N$ small, after $m = \frac\pi4 \sqrt{N/\ell}$ iterations, 
the probability of reaching $|\wt w\rangle$ is close to 1.$\hfill\square$
\end{cor}

The result (\ref{grov:eq3.12}) is consistent with Grover's original algorithm 
for single object search with $\ell=1$, which has $m\approx \frac\pi4 \sqrt N$.

\begin{thm}\label{grov:thm3.6} {\bf (Boyer, Brassard, H\o yer and Tapp 
[\ref{BBHT}]).}
Assume that $\ell/N$ is small. Then any search algorithm for $\ell$ objects, in 
the form of
$$U_pU_{p-1}\ldots U_1|w_I\rangle,$$
where each $U_j, j=1,2,\ldots, p$, is a unitary operator and $|w_I\rangle$ is an 
arbitrary combination state, takes in average $p=\mathcal{O}(\sqrt{N/\ell})$ 
iterations in order to reach the subspace $L$ with a positive probability $P$ 
independent of $N$ and $\ell$. 
Therefore, the generalized Grover algorithm in Cor.~\ref{grov:cor3.5} is of 
optimal order.
\end{thm}

\begin{proof}
This is the major theorem in [\ref{BBHT}]; see Section~7 and particularly
Theorem~8 
therein.
Note also the work by C.~Zalka who considered some measurement effects in 
[\ref{zalka}].
\end{proof}

Unfortunately, if the number $\ell$ of good items is not known in 
advance, Corollary~\ref{grov:cor3.5} does not tell us when to stop 
the iteration.
 This problem was addressed in~[\ref{BBHT}], and in another way 
 in~[\ref{BHT}].
 In a related context an equation arose that was not 
fully solved in~[\ref{BBHT}].
We consider it in the final segment of this section.
   As in [\ref{BBHT},~\S 3], consider stopping the Grover process 
   after $j$ iterations, and, if a good object is not obtained, 
 starting it over again from the beginning.
  From Corollary~\ref{grov:cor3.5}, the probability of success 
  after $j$ iterations is $\cos^2(j\theta -\alpha)$.
 By a well-known theorem of probability theory, if the probability 
of success in one ``trial'' is~$p$,
 then the expected number of trials before success is achieved 
will be $1/p$.
(The probability that success is achieved on the $k$th trial
is $p(1-p)^{k-1}$.
Therefore, the expected number of trials is
\begin{equation}
\sum_{k=1}^\infty kp(1-p)^{k-1} =
-p \sum_{k=1}^\infty \frac d{dp} (1-p)^k =
-p\, \frac d{dp} \frac{1-p}p \;,
\end{equation}
which is $1/p$.)
 In our case, each trial consists of $j$ Grover iterations, so the 
expected number of iterations before success is
$$E(j) = j\cdot \sec^2 (j\theta-\alpha)\;.$$  
The optimal number of iterations $j$ is obtained by
 setting the derivative $E'(j)$ equal to zero: 
\begin{align}
0 = E'(j) &= \sec^2(j\theta-\alpha)
  +2j\theta\sec^2(j\theta-\alpha)  
\tan(j\theta-\alpha),\nonumber\\ 
\label{grov:eq3.13}
2j\theta &= -\cot((j\theta-\alpha))\;. 
\end{align}
 (In [\ref{BBHT}, \S 3], 
 this 
equation is derived 
 in the form $4\vartheta j = \tan((2j+1)\vartheta)$, 
 which is seen to be equivalent to (\ref{grov:eq3.13}) by noting 
 that $\vartheta = \frac{\theta}2 = \frac{\pi}2 - \alpha$. 
Those authors then note that they have not solved the equation 
 $4\vartheta j = \tan((2j+1)\vartheta)$ 
 but proceed to use an ad hoc equation $z = \tan(z/2)$ with 
$z=4\vartheta j$ instead.) 
 Let us now approximate the solution $j$ of 
(\ref{grov:eq3.13}) 
iteratively as follows. From (\ref{grov:eq3.13}),
\begin{align}
&2j\theta \sin(j\theta-\alpha) + \cos(j\theta-\alpha)
  = 0\;,\nonumber\\
\label{grov:eq3.14}
&e^{2i(\theta j-\alpha)} = (i2\theta j+1)/(i2\theta j-1)\;, 
\end{align}
and by taking the logarithm of both sides, we obtain
\begin{equation}\label{grov:eq3.15}
2i(\theta j-\alpha) = 2i\pi n 
 + i\arg\left(\frac{i2\theta j+1}{i2\theta 
j-1}\right) + \ln\left|\frac{i2\theta j+1}{i2\theta j-1}\right|\;,
\end{equation}
for any integer $n$. Assume that $\ell/N$ is small so that $j$ is large, but we 
are looking for the smallest such positive $j$. Note that the logarithmic term 
in (\ref{grov:eq3.15}) vanishes, and
\begin{align*}
\arg\left(\frac{i2\theta j+1}{i2\theta j-1}\right) &=
 - 2\tan^{-1} \frac1{2\theta 
j}\\
&= 2\left[\sum^\infty_{q=0} \frac{(-1)^{q+1}}{2q+1}
  \left(\frac1{2\theta 
j}\right)^{2q+1}\right]\\
&= -\, \frac1{\theta j} + \mathcal{O}((\theta j)^{-3})\;; 
\end{align*}
by taking $n=0$ in (\ref{grov:eq3.15}), we obtain
\begin{align}
j &= \frac1{2i\theta} \left[2i\alpha 
  - i\cdot \frac1{\theta j} + 
\mathcal{O} ((\theta j)^{-3})\right]\nonumber\\
\label{grov:eq3.16}
&= \frac1\theta \left[\alpha  - \frac1{2\theta j} + 
\mathcal{O}((\theta j)^{-3})\right]\;.
\end{align}

The first order approximation $j_1$ for $j$ is obtained by solving
\begin{align}
j_1 &= \frac1\theta \left(\alpha
  - \frac1{2\theta j_1}\right)\;, 
\nonumber\\
j^2_1 &- \frac1\theta\,  \alpha j_1
 + \frac1{2\theta^2} =0\;, 
\nonumber\\
\label{grov:eq3.17}
j_1 &= \frac1{2\theta} (\alpha + 
\sqrt{\alpha^2-2})\;. 
\end{align}
Higher order approximations $j_{n+1}$ for $n=1,2,\ldots,$ may be obtained by 
successive iterations
$$j_{n+1} = \frac1\theta \left(\alpha  - \tan^{-1} 
\frac1{2\theta j_n}\right)$$
based on (\ref{grov:eq3.13}). This process will yield a convergent 
solution $j$ to (\ref{grov:eq3.13}).

\section*{Appendix: Random Multi-Object Search}

\renewcommand{\theequation}{A.\arabic{equation}}
\renewcommand{\theprop}{A.\arabic{prop}}
\renewcommand{\thethm}{A.\arabic{thm}}
\renewcommand{\therem}{A.\arabic{rem}}

\setcounter{equation}{0}
\setcounter{prop}{0}

\indent

Given a set of $N$ unsorted objects, among which $\ell$ of them are the
desired objects that we are searching for, how many times in average do we
need to search in order to obtain the first desired object?

Because the $N$ objects are unsorted, the above problem is equivalent to a
familiar problem in probability theory:

\begin{align}
&\text{``An urn contains $N$ balls, $\ell$ of them are black and the rest are
white. Each time}\nonumber\\
&\text{draw a ball randomly without replacement. Determine the number of times
in}\nonumber\\
\label{eqA.1}
&\text{average needed in order to draw the first black
ball''.} 
\end{align}
A different version of random multi-object search would correspond to
(\ref{eqA.1}) but {\em with replacement\/} after each  drawing. This search
method is less efficient than (\ref{eqA.1}), but can be treated by a similar
approach as given below and, thus, will be omitted.

Even though we believe the solution to (\ref{eqA.1}) is available in the
literature, we could not locate a precise citation and, therefore, feel the
urge to write this Appendix.

We define a random variable
\begin{align}
T_b \equiv &\text{ number of drawings needed to draw a  ball randomly}
\nonumber\\
\label{eqA.2}
\phantom{equiv} &\text{ and without replacement until the first black ball is
out.}
\end{align}
Our objective is to calculate $E(T_b)$, the expected or the average value of
$T_b$.

Obviously,
\begin{equation}\label{eqA.3}
T_b\in \{1,2,\ldots, N-\ell+1\}.
\end{equation}
We use $\binom{n}{j}$ to denote the combinatorial coefficient $n!/[j!(n-j)!]$,
and use $P(A)$ to denote the probability of a given event $A$ (measurable with
respect to the random variable $T_b$).

\begin{prop}\label{proA.1}
For $j\in \{1,2,\ldots, N-\ell+1\}$,
\begin{equation}\label{eqA.4}
P(T_b=j) = \frac{\binom{N-\ell}{j-1}}{\binom{N}{j-1}} \cdot \frac\ell{N-(j+1)}.
\end{equation}
\end{prop}

\begin{proof}
By the very definition of $T_b$ in (\ref{eqA.2}), we know that
\begin{align*}
P(T_b = j) &= P(\text{the first $j-1$ drawings result in $j-1$ white balls,
but}\\
&\phantom{=} \text{the $j$-th drawing results in a black ball).}
\end{align*}
Therefore (\ref{eqA.4}) follows.
\end{proof}

\begin{prop}\label{proA.2}
([\ref{Kn}, p.~54, (10), (11)]) \ For any $m,n\in \{0,1,2,\ldots\}$ and $m\le
n$, 
$$\sum^{n-m}_{k=0} \binom{m+k}m = \sum^n_{j=m} \binom{j}m = \binom{n+1}{m+1}$$
\end{prop}

\begin{proof}
By Pascal's formula,
\begin{equation}\label{eqA.5}
\binom{j+1}{m+1} = \binom{j}{m+1} + \binom{j}m,
\end{equation}
we have
\begin{align*}
\sum^n_{j=m} \binom{j}m &= \sum^n_{j=m} \left[\binom{j+1}{m+1} -
\binom{j}{m+1} \right]\\
&= \sum^n_{j=m} \binom{j+1}{m+1} - \sum^{n-1}_{j=m-1} \binom{j+1}{m+1}\\
&= \binom{n+1}{m+1} - \binom{m}{m+1} = \binom{n+1}{m+1}.\qed
\end{align*}
\renewcommand{\qed}{}\end{proof}

\begin{thm}\label{thmA.3}
\begin{equation}\label{eqA.6}
E(T_b) = \frac{N+1}{\ell+1}.
\end{equation}
\end{thm}

\begin{proof}
From (\ref{eqA.4}), we have
\begin{align}
E(T_b) &= \sum^{N-\ell+1}_{j=1} j\cdot \frac{\binom{N-\ell}{j-1} \cdot\ell}{
\binom{N}{j-1} \cdot [N-(j-1)]} = \sum^{N-\ell+1}_{j=1} j\cdot\ell \cdot
\dfrac{\frac{(N-\ell)!}{(j-1)! (N-\ell-j+1)!}}{\frac{N!}{(j-1)! (N-j+1)!}
\cdot (N-j+1)}\nonumber\\
&= \frac{\ell(N-\ell)!}{N!} \sum^{N-\ell}_{j=0} (j+1) \dfrac{(N-j-1)!}{(N -
\ell-j)!}\nonumber\\
&= \dfrac{\ell(N-\ell)!}{N!} \sum^{N-\ell}_{k=0} (N-\ell+1-k) \cdot
\dfrac{(k+\ell-1)!}{k!} \quad \text{(where $k=N-\ell-j$)}\nonumber\\
&= (N-\ell+1) \left[\dfrac{\ell(N-\ell)!}{N!} \sum^{N-\ell}_{k=0}
\dfrac{(k+\ell-1)!}{k!}\right] - \dfrac{\ell(N-\ell)!}{N!} \sum^{N-\ell}_{k=1}
\dfrac{(k+\ell-1)!}{(k-1)!}\nonumber\\
&= (N-\ell+1) \left[\sum^{N-\ell}_{k=0}
\dfrac{\frac{(k+\ell-1)!}{k!(\ell-1)!}}{\frac{N!}{\ell!(N-\ell)!}}\right] -
\ell\left[ \sum^{N-(\ell+1)}_{k=0}
\dfrac{\frac{(k+\ell)!}{k!\ell!}}{\frac{N!}{\ell!(N-\ell)!}}\right]\nonumber\\
\label{eqA.7}
&= (N-\ell+1) \left[\frac1{\binom{N}\ell} \sum^{N-\ell}_{k=0}
\binom{k+\ell-1}{\ell-1}\right] - \ell \left[\frac1{\binom{N}\ell}
\sum^{N-(\ell+1)}_{k=0} \binom{k+\ell}\ell\right].
\end{align}
Now, applying Proposition~A.2,  we obtain
\begin{align}
\label{eqA.8}
\binom{N}{\ell-1} + \sum^{N-\ell}_{k=0} \binom{k+\ell-1}{\ell-1} &=
\binom{N}\ell ,\\ 
\label{eqA.9}
\binom{N}\ell + \sum^{N-(\ell+1)}_{k=0} \binom{k+\ell}\ell &= \binom{N}{\ell+1}
 = \dfrac{N-\ell}{\ell+1} \binom{N}\ell.
\end{align}
Substituting (\ref{eqA.8}) and (\ref{eqA.9}) into (\ref{eqA.7}), we obtain
\begin{align*}
E(T_b) &= \cdots \text{(continuing from (\ref{eqA.7}))}\\
&= (N-\ell+1) \binom{N}{\ell}^{-1} \binom{N}{\ell-1}-\ell + (N-\ell+1) - \ell
\cdot \frac{N-\ell}{\ell+1} = \frac{N+1}{\ell+1}.\qed 
\end{align*}
\renewcommand{\qed}{}\end{proof}

\begin{rem}\label{remA.1}
When $\ell=1$, by (\ref{eqA.6}) it takes $(N+1)/2$ searches in average to
obtain the desired single object. In the literature, this is usually cited as
 $N/2$, which  of course differs negligibly for large $N$.

For $N=4$ and $\ell=1$, by (\ref{eqA.6}) it takes $(4+1)/(1+1) = 2.5$ times of
search on average to obtain the desired item. But in [\ref{Chu}, p.~3408], it
is stated that it takes $9/4 =2.25$  times of search on average.
The reason for the discrepancy is a different definition of successful search.
The authors of [\ref{Chu}] regard the search as completed as soon as the
location of the desired item is known, even if that item has not been
physically ``drawn from the urn''. They therefore count the worst case, where
the desired item is the last one drawn, as requiring only 3 steps instead of
4. This redefinition could be incorporated into our theorem at the expense of
some complication; but it seems to us to be the less natural convention in the
scenario of multiple desired objects only one of which is required to be
``produced''.

$\hfill\square$
\end{rem}

\medskip
{\bf Acknowledgments:} 
We thank M.O.~Scully for originally expounding [\ref{FG}], B.-G. Englert and M.
Hillery for acquainting us with some of the literature of quantum computation, 
M.~M.~Kash for a technical discussion, and Hwang~Lee, D.~A.~Lidar, and
J.~D.~Malley for comments on the manuscript.

\end{document}